\documentclass{iau} 
\usepackage{graphicx}

\title[Light-curves of unidentified sources] 
{The Study of Variability  of 8 Blazar Candidates Among the \textit{Fermi}-LAT Unidentified Gamma-Ray Sources}

\author[Nkundabakura P. et al]   
{Nkundabakura P.$^{1},$
Kamanzi  J. D'amour$^{2},$
Mbarubucyeye J. D.$^{1,3},$
and Mutabazi T.$^{2}$ }

\affiliation{$^{1}$University of Rwanda, College of Education, P.O.  Box 5039, Kigali, Rwanda \\[\affilskip] 
$^{2}$Mbarara University of Science and Technology, Department of Physics, Mbarara, Uganda  \\[\affilskip]
$^{3} $Deutsches Elektronen-Synchrotron (DESY), Platanenallee 6, 15738 Zeuthen, Germany.
}

\pubyear{2019}
\volume{356}  
\jname{Nuclear Activity in Galaxies Across Cosmic Time}
\editors{M. Povic, J.Masegosa, H. Netzer, P. Marziani, \\P. Shastri,
S.B. Tessema \& S. H. Negu, eds.}
\begin{document}

\maketitle

\begin{abstract}
We discuss the  time-series behavior of 8 extragalactic 3FGL sources away from the Galactic plane (i.e., $\mid b\mid \geq 10^{\circ}$) whose uncertainty ellipse contains a single X-ray and one radio source. The analysis was done using the standard Fermi \textit{ScienceTools}, package of version v10r0p5. The results show that sources in the study sample display a slight indication of flux variability in $\gamma$-ray on monthly timescale. Furthermore, based on the object location on the variability index versus spectral index diagram, the positions of 4 objects in the sample were found to fall in the region of the already known BL Lac positions.
\end{abstract}

\firstsection 
\section{Introduction}

\label{intro}
\noindent The study of variability is particularly important in $\gamma$-ray astronomy primarily due to different advantages such as assisting in the identification of the correct radio/optical/X-ray source within the $\gamma$-ray position box, with the observations at other wavelengths (\cite[De Cicco et al. 2015]{De2015}; \cite[Ferrara et al. 2015]{ferrara2015}). For unidentified sources, variability characteristics can also support the recognition of the correct source class (\cite[Nolan et al. 2003]{ Nolan2003}).\par

\noindent Fortunately, the Large Area Telescope (LAT) aboard the \textit{Fermi Gamma-ray Space Telescope} has revolutionised the field of $\gamma$-ray astronomy by detecting a wealth of new $\gamma$-ray sources and allowing the study of previously known sources with unprecedented details (\cite[Zechlin \& Horns. 2015]{Zechlin2015}). Previous studies show that most of the sources detected by the \textit{Fermi}-LAT are blazars (\cite[Ackermann et al. 2015]{Ackermann2015}).  The 3FGL~(\cite[Acero et al. 2015]{Acero}) and the 4FGL~(\cite[The Fermi-LAT collaboration 2019]{Collabo2019}) catalogs reported a significantly large fraction of sources compared to the previous ones.

\noindent However, the majority of 3FGL and 4FGL sources remain unassociated  with low-energy counterparts, hence understanding their nature is an open question in high-energy astrophysics. In addition, it seems plausible that most of the unassociated high-latitude $\gamma$-ray sources are expected to be faint AGN, which may include blazar sub-class (\cite[Mirabal et al. 2012]{2012Mirabal};~\cite[Massaro et al. 2012]{Massaro}; \cite[Ackermann et al. 2012]{Ackermann}). These unidentified $\gamma$-ray sources represent a discovery area for the new source classes or new members of existing source classes which may include different types of AGN.\par

\noindent For instance, previous studies show a combined effort to isolate potential blazar candidates among this large population (e.g.~\cite[Massaro et al. 2012]{Massaro}; \cite[Zechlin \& Horns  2015]{Zechlin2015}; \cite[Paiano et al. 2017]{2017Paiano}). Some studies used the analysis of the  multiwavelength Spectral Energy Distribution (SED) through detecting a double peaked spectrum. This indicated that the radiation among the selected sample originates mainly from synchrotron and the inverse-Compton emission in the so-called synchrotron-Compton blazars (\cite[Mbarubucyeye J.D.,  Krau\ss ~F.,  and Nkundabakura P.  2019 in prep...]{Mbarubucyeye}), though the SED alone is not enough to fully characterise the blazar nature based on their broad band properties. Since  blazars display intrinsic variability and more significantly in the $\gamma$-ray energy band (\cite[Ulrich et al. 1997]{Ulrich}), it is needed to use this property to characterise individual synchrotron-Compton blazar candidates that may be present in the $\gamma$-ray unidentified and unassociated population. 
In this paper, we discuss the  time-series behavior of 8 extragalactic 3FGL sources away from the Galactic plane (i.e., $\mid b\mid \geq 10^{\circ}$) which were carefully selected among the Unidentified \textit{Fermi}-LAT sources with the purpose  to detect any sign of  variability which can be linked  to the blazar nature of these sources.\\

\section{Sample selection}
\label{sect.2}
The following selection criteria were used to obtain a study sample:
\begin{enumerate}[i.]
 \item Being unidentified sources at high Galactic latitudes, $\mid b\mid \geq 10^{\circ}$,
 \item Being unidentified sources which have a single X-ray and one radio source in its uncertainty region,
 \item Being unidentified sources that were reported in the 4FGL catalog.
\end{enumerate}
Applying all cuts to the population of unidentified sources listed in 3FGL, a sample of 8 unidentified sources thought to be potential blazar candidates was isolated.

\section{Data analysis}
\label{sect.4}
\noindent The astrophysical data analysis of LAT  begins with a list of counts detected. This list results from processing made by the LAT instrument team, which reconstructs events for the signals from different parts of LAT. Two principal types of analysis were applied in this study, they were performed in a systematic way such that the results from the first analysis became the input of the next one. The types of analysis performed are:
\begin{itemize}
 \item Global analysis which was performed using \textit{Fermi} \textit{ScienceTools} v10r0p5. This provided the fluxes and spectral parameters of all objects in our study sample. The photon counts within a region of interest of 25 degree radius were taken into account. We selected events within the energy range 100 MeV--300 GeV, a maximum zenith angle of 90 degrees and event type 3.
 \item Time-series analysis (light-curve analysis \& variability analysis). This provided the $\gamma$-ray light-curves for the period of 9 years and the variability indices of target sources, together with the significance of the observed variability in light-curves. 
\end{itemize}

\noindent The observed variability was obtained using the following equation as in \cite[Nolan et al. 2003]{ Nolan2003}:
\begin{equation}\label{eq:0}
 TS_{var}=2\sum_{i} \frac{\Delta F_i^2}{\Delta F_i^2 +f^2 F_{const}^2} \ln\left(\frac{ \mathcal{L}_{i}(F_{i})}{\mathcal{L}_{i}(F_{const})}\right), 
\end{equation}
where $f = 0.02$, i.e., a $2\% $, is a systematic correction factor, $F_{i}$ and $\Delta F_i$ are the flux and error in flux in the $i^{th}$ bin, respectively. $\mathcal{L}_{i}(F_{const})$ is the value of the likelihood in the $i^{th}$ bin under the null hypothesis where the source flux is constant across the full period and $F_{const}$ is the constant flux for this hypothesis.  $\mathcal{L}_{i}(F_{i})$  is the value of the likelihood in the $i^{th}$ bin under the alternate hypothesis where the flux in the $i^{th}$ bin is optimised.

\section{Results and discussion}
\label{sect.5}
\subsection{Average results}
Before performing the light-curve analysis, a fit of the entire 108 months LAT data using a power-law model was performed through binned likelihood analysis for each target  source. This provided the sample average results fluxes and spectral properties in a period of 9 years. Sources in the sample were found to be faint $\gamma$-ray emitters with $\gamma$-ray spectral index, $\Gamma \sim 2$ (see Table ~\ref{Global}), which is consistent with the previous studies (e.g., \cite[Ackermann et al.  (2015)]{Ackermann2015}).
\begin{table}
 \centering
    \caption{Gamma-Ray average fluxes and spectral characteristics of the sample sources. }
    \label{Global}
    \scriptsize
    \setlength{\tabcolsep}{2pt}
    \begin{tabular}{lccccccr} 
      No&3FGL Name& \hspace{2mm} $\centering{F_{\gamma}}$& \hspace{2mm} $\centering{\Gamma_{\gamma}}$&\hspace{2mm} $\sigma$&\\
                 &(1) &(2) &(3)&(4)\\
      \hline\hline
     1 & J0049.0+4224 &\hspace{2mm} 0.93 $\pm$ 0.43&\hspace{2mm} 1.81 $\pm$ 0.14&\hspace{2mm} 7.02\\
     2 & J1119.8$-$2647&\hspace{2mm} 2.99 $\pm$ 0.81&\hspace{2mm} 1.94 $\pm$ 0.11&\hspace{2mm} 10.03\\
     3 & J1132.0$-$4736&\hspace{2mm} 3.72 $\pm$ 1.21&\hspace{2mm} 2.00 $\pm$ 0.09&\hspace{2mm} 10.11\\
     4 & J1220.0$-$2502&\hspace{2mm} 7.62 $\pm$ 1.26&\hspace{2mm} 2.16 $\pm$ 0.20&\hspace{2mm} 7.46\\
     5 & J1220.1$-$3715&\hspace{2mm} 3.44 $\pm$ 0.81&\hspace{2mm}1.96 $\pm$ 0.09&\hspace{2mm} 10.24\\
     6 & J1619.1+7538&\hspace{2mm} 0.85 $\pm$ 0.23&\hspace{2mm} 1.78 $\pm$ 0.10&\hspace{2mm} 10.01\\
     7 & J1923.2$-$7452&\hspace{2mm} 8.61 $\pm$ 0.68&\hspace{2mm} 2.04 $\pm$ 0.10&\hspace{2mm} 14.05\\
     8 & J2015.3$-$1431&\hspace{2mm} 4.63 $\pm$ 1.52&\hspace{2mm} 2.23 $\pm$ 0.19&\hspace{2mm} 5.03\\ \hline \hline
    \end{tabular}
    \\ Note: Column 1, 2, 3 and 4 show the source 3FGL name, the average $\gamma$-ray flux for 108 months LAT data in scale of $10^{-9} \, \rm{ph} \, \rm{cm^{-2}} \, \rm{s^{-1}}$, the $\gamma$-ray spectral index corresponding to column (2), the significance in sigma units corresponding to column (2), respectively.
\end{table}
\subsection{Monthly $\gamma$-ray light-curves}
To determine the trends of flux change and variability of sources for a period of 9 years, light-curve analysis was performed. This was done through extracting the monthly fluxes along this period, and plotting light-curves.\\

\begin{figure}
\centering
 \includegraphics[width=9.2cm,height=6cm,angle=0]{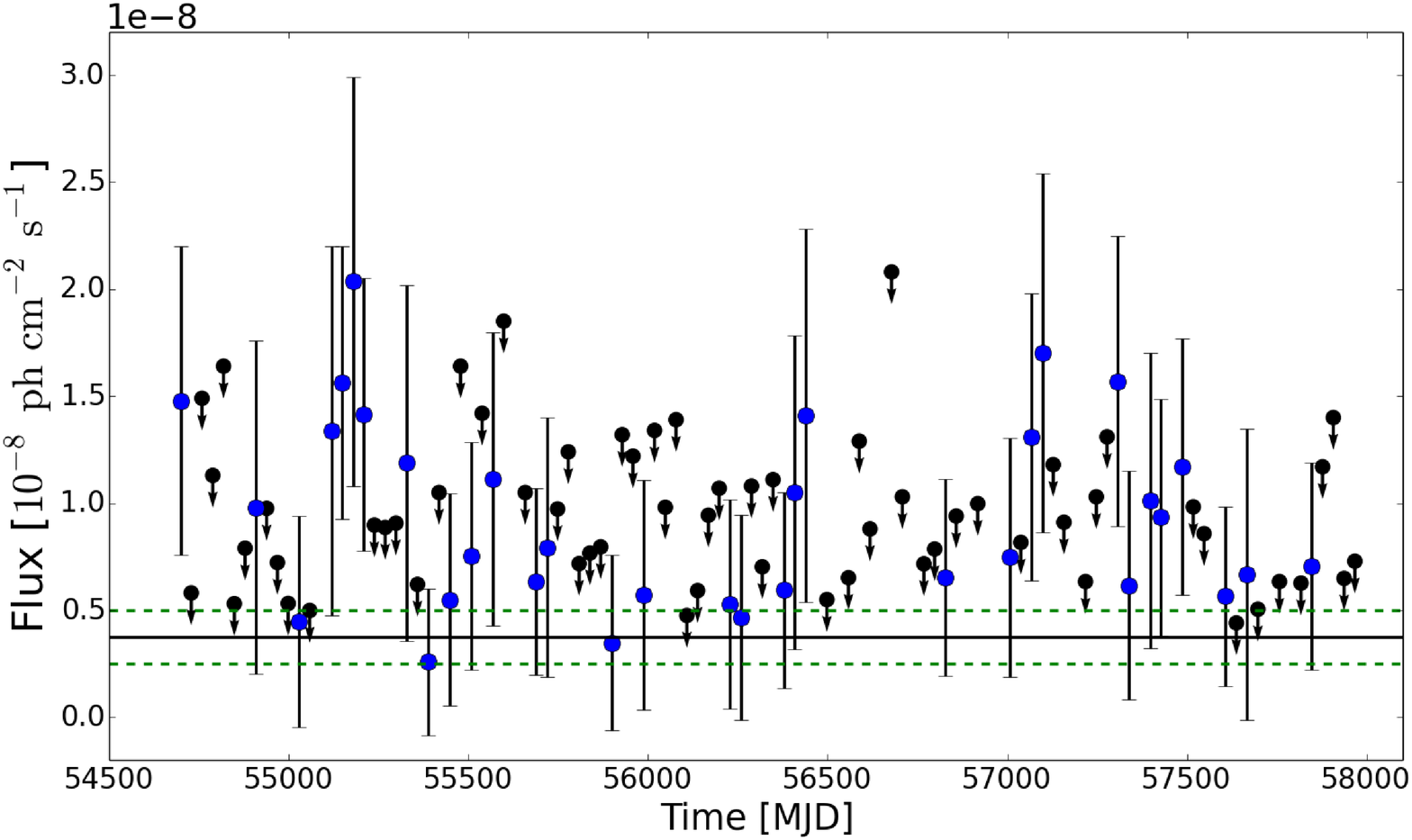}
 \caption[The monthly $\gamma$-ray light-curves for sources in the study sample]{The 100 MeV--300 GeV monthly light-curves for 3FGL J1132.0$-$4736. The horizontal solid line along with two dashed lines present the 9-years average flux and its 1$\sigma$ error range derived in the global analysis, respectively. The blue points represent flux with its $1\sigma$ error bar, while the downward arrows together with black points represent the 95\% upper limits.}
\label{lch1}
\end{figure}

\noindent Generally, we found that sources in our sample do not show significant signal in their light-curves, which is an indication that they are relatively faint in $\gamma$-ray. This was also suggested by \cite[Acero et al.  (2015)]{Acero}. Therefore, the light-curves indicate that signals are not significantly detectable in many monthly bins (represented as an upper limit). This implies  that their fluxes are close to zero, hence summing them over the full period (9 years) tends to lower the source average flux as shown in Figure \ref{lch1}. The sample light-curves display behaviours commonly shown by blazars such as: non periodic flux change characterised by undefined and no specific trends, associated with unpredictable and sudden flux rise seen across the whole period of 9 years (see Figure \ref{lch1}). However, the large error bars on the data points does not allow to firmly establish such sharp flux rises. 

\subsection{Variability indices}
\label{vari}
\begin{figure}
\centering
 \includegraphics[width=0.6\linewidth]{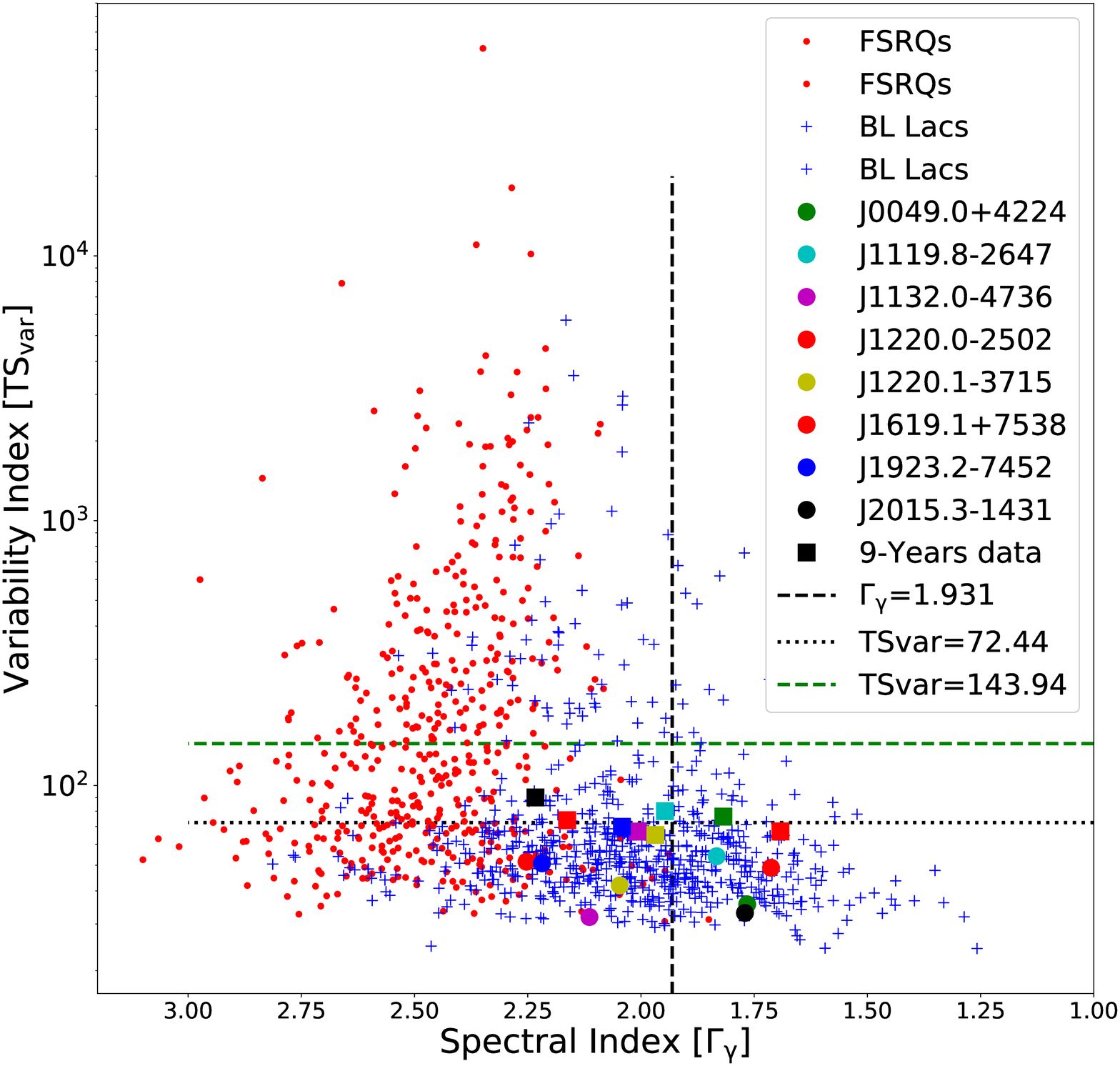}
\caption[The variability index versus spectral index diagram: Extragalactic sources]{Variability index ($TS_{var}$) versus power law spectral index (PL index) diagram of all 3FGL extragalactic sources. Red points: Flat Spectrum Radio Quasars, blue crosses: BL Lacs, and sources in the study sample. Circles in different colours indicate data of sources in the selected sample, while solid squares in same colours present results of the same sources obtained in this study. The black horizontal dotted line indicates the 3FGL variability index threshold (72.44), while the green horizontal dashed line shows the 9-years variability index threshold (143.94) obtained in this study. The black vertical dashed line shows the spectral index (1.931) below which the region is populated by BL Lacs.}
\label{TSvarE}
\end{figure}

The light-curves presented in this study show  many upper limits that correspond to the time when the signal in monthly bins was not significant enough to characterise a source. It is also clear that the error bars corresponding to the significant flux points are relatively large. Therefore, we used the `\textit{variability index}' defined in Equation ~\ref{eq:0} to quantify the observed variability in light-curves, in which the information of the upper limits is properly considered.
To compare the already known classification in 3FGL  with our sources that are lacking classifications, variability-spectral index diagram for all 3FGL sources  including the study sample was plotted.
\cite[Ackermann et al. (2015)]{Ackermann2015} observed that blazars are located in different zones on the variability-spectral index diagram, according to their subtypes (FSRQs and BL Lacs), though there is also a large recovery zone (see Figure \ref{TSvarE}).

The significance of the observed variability from sample light-curves was estimated by using a $\chi^2$ distribution. This provided the variability index ($TS_{var}$) threshold at which we assigned the source a 99\% probability of being variable (on a timescale of $\gtrsim 1 \, \rm{month}$). For 9-years data, we found that variability is considered significant with 99\% confidence level if the variability index is greater than 143.94. However, sources in our study sample have $TS_{var}$ values much lower than the threshold (i.e., $TS_{var} < 143.94$). This implies that we can not conclude at 99\% confidence level that our target sources are variable due to lack of statistics. The variability significance of all sources in the study sample was found to be in the range of 0.5\% to 12\%. The variability significance of 3FGL J2015.3-1431 was found to be the highest compared to other sources in the study sample. 
\section{Conclusions}
\label{sec.6}
\noindent Although probing the $\gamma$-ray variability of blazar candidate sources is of definite interest in the study of AGN properties towards a better classification of the sources, definite classification is expected to be properly achieved by multiwavelength studies of  their spectral energy distribution together with their optical spectra. Indeed, variability is well understood when it is studied across the electromagnetic spectrum (Radio, Optical/UV and X-rays), and on different timescales. This contributes to checking the variability correlation in different energy bands and testing whether variability exists for all timescales. Therefore, future studies are expected to consider multi-waveband variability and on different timescales. The Variability can be applied to estimate physical parameters of AGN such as the size of the emitting region, timescale of variability, magnetic field in the jets, mass of the central engine (blackhole), etc. However, the estimation of all these parameters requires primarily to know the object's redshift, which can be obtained through spectroscopic studies. Therefore, future studies through the analysis of optical spectra of sources listed in our studied sample should be considered. Such observations from ground-based optical telescopes (such as a 10-meter class telescope) would be the ideal program to determine the nature of blazar candidates.
\section{Acknowledgements}
\label{sec.7}
\noindent We acknowledge the useful contribution of Richard J.G. Britto, University of the Free State - South Africa. Financial support from the Swedish International Development Cooperation Agency (SIDA) through the International Science  Programme (ISP)  is also  gratefully acknowledged.

\end{document}